\documentclass[
  twocolumn,
  prl,
  showpacs,
  amsmath,
  amssymb,
  superscriptaddress,
  floatfix
]{revtex4-1}

\usepackage{bm}
\usepackage{graphicx}
\usepackage{multirow}
\usepackage[dvips]{color}
\usepackage{enumerate}

\newcommand{\be}{\begin{equation}}
\newcommand{\ee}{\end{equation}}
\newcommand{\bea}{\begin{eqnarray}}
\newcommand{\eea}{\end{eqnarray}}
\newcommand{\ba}{\begin{array}}
\newcommand{\ea}{\end{array}}

\newcommand{\LMon}{\mathcal{L}} 
\newcommand{\RMon}{\mathcal{R}} 

\newcommand{\minlr}{\mathfrak{m}} 
\newcommand{\galr}{\gamma^{\LMon,\RMon}}
\newcommand{\honelr}{h^{\LMon,\RMon}(1)}
\newcommand{\galrL}{\gamma_L^{\LMon,\RMon}}
\newcommand{\honelrL}{h_L^{\LMon,\RMon}(1)}

\newcommand{\up}{\uparrow}
\newcommand{\down}{\downarrow}

\usepackage{tikz}
\usetikzlibrary{shapes.geometric}
\usetikzlibrary{decorations.markings}
\usepgflibrary{decorations.shapes}
\usepgflibrary{shapes.misc}

\begin{document}

\title{Exact exponents for the spin quantum Hall transition in the presence of multiple edge channels}

\author{R.~Bondesan}
\affiliation{Institute de Physique Th\'eorique, CEA Saclay, F-91191
  Gif-sur-Yvette, France}
\affiliation{LPTENS, \'Ecole Normale Sup\'erieure, 24 rue Lhomond,
  75231 Paris, France}
\affiliation{Institut Henri Poincar\'e, 11 rue Pierre et Marie Curie,
  75231 Paris, France}

\author{I.~A.~Gruzberg}
\affiliation{Institut Henri Poincar\'e, 11 rue Pierre et Marie Curie,
  75231 Paris, France}
\affiliation{James Franck Institute, University of Chicago, 5640 South Ellis
Avenue, Chicago, Illinois 60637, USA}

\author{J.~L.~Jacobsen}
\affiliation{LPTENS, \'Ecole Normale Sup\'erieure, 24 rue Lhomond,
  75231 Paris, France}
\affiliation{Institut Henri Poincar\'e, 11 rue Pierre et Marie Curie,
  75231 Paris, France}
\affiliation{Universit\'e Pierre et Marie Curie, 4 place Jussieu,
  75252 Paris, France}

\author{H.~Obuse}
\affiliation{Institut Henri Poincar\'e, 11 rue Pierre et Marie Curie,
  75231 Paris, France}
\affiliation{Department of Physics, Kyoto University, Sakyo-ku, Kyoto 606-8502, Japan}
\affiliation{Institut f\"ur Nanotechnologie, Karlsruhe Institute of Technology, 76021 Karlsruhe, Germany}

\author{H.~Saleur}
\affiliation{Institute de Physique Th\'eorique, CEA Saclay, F-91191 Gif-sur-Yvette, France}
\affiliation{Institut Henri Poincar\'e, 11 rue Pierre et Marie Curie,
  75231 Paris, France}
\affiliation{Physics Department, USC, Los Angeles, CA 90089-0484, USA}

\begin{abstract}

Critical properties of quantum Hall systems are affected by the presence of extra edge channels---present, in particular, at higher plateau transitions.  We study this phenomenon for the case of the spin quantum Hall transition. Using supersymmetry we map the corresponding network model to a {\it classical} loop model, whose boundary critical behavior was recently determined exactly. We verify predictions of the exact solution by extensive numerical simulations.

\end{abstract}


\maketitle

More than fifty years after its discovery, Anderson localization \cite{AL50} remains a vibrant research field. One central research direction is the physics of Anderson transitions (AT) \cite{evers08}, including metal-insulator and quantum Hall (QH) type transitions (that is, between different phases of topological insulators). Apart from electronic gases in semiconductor structures, experimental realizations include localization of light \cite{wiersma97}, cold atoms \cite{BEC-localization}, ultrasound \cite{faez09}, and optically driven atomic systems \cite{lemarie10}. Theoretically, the field was strongly boosted by the discovery of unconventional symmetry classes and a complete symmetry classification of disordered systems \cite{altland97, zirnbauer96, evers08, heinzner05}.

Recently it was realized that AT in systems with boundaries may exhibit boundary critical behavior different from, and richer than, the bulk behavior \cite{subramaniam06, mildenberger07, obuse08a}. The boundary criticality has served to test conformal invariance at two-dimensional AT \cite{conf-inv, obuse10, ryu10} and as a strong constraint on possible theories of the integer QH transition \cite{qhe}. These works often employed the so-called network models \cite{networks,kramer05} of AT for numerical studies. Within a network formulation the richness of boundary critical behaviors relates to the possibility of having multiple edge channels at the boundary \cite{obuse08a, Barnes93, Kobayashi09}. Physically,  multiple edge channels occur in an integer QH system whenever the filling fraction exceeds one.  Some of the results below may directly apply to  the physics of higher QH plateaus and transitions between them. 
However, our description neglects electron interactions at the edge
which could be relevant in experimental realizations of the QH effect
\cite{Chklovskii92,Silvestrov08}.

In this Letter we study boundary critical properties in the presence of multiple edge states at the so-called spin quantum Hall (SQH) transition \cite{senthil99}. The corresponding network model \cite{senthil99, kagalovsky99} enjoys a very special status. In the bulk, or with reflecting boundaries, the model in its minimal formulation (suitable to describe mean conductances) can be mapped to classical percolation on a square lattice \cite{gruzberg99, subramaniam08}. This mapping determines exact critical properties at the SQH transition \cite{gruzberg99, cardy00, beamond02, subramaniam08, mirlin03}. In this Letter we demonstrate that extra edge channels can be straightforwardly included in the mapping. The resulting classical model is not percolation any more, but can nonetheless be formulated as a loop model.

Both network and loop models (or percolation) are lattice regularizations of field-theoretic descriptions of AT in terms of sigma-models on symmetric superspaces \cite{SUSY, zirnbauer96, evers08}. This connection is thoroughly explained in Ref.~\cite{Read2001}, and was recently extensively explored by some of us \cite{saleur_et_al}. Through this connection, complete spectra of boundary operators were obtained for the conformal sigma-models on superspaces $\mathbb{CP}^{N+M-1|N}$ with a topological theta-angle \cite{bondesan11}. In the sigma-model approach, the number of extra edges is related to the exact value of the theta-angle, which affects boundary (but not bulk) properties \cite{xiong97}. The case $N=M=1$ is directly relevant to the SQH effect, and we here apply results of Ref.~\cite{bondesan11} to obtain exact exponents describing scaling of the mean boundary point contact conductances in the presence of multiple edges.

We also report extensive numerical simulations of mean conductances in network models on open strips with edge channels on both sides. Conformal invariance (which has been numerically demonstrated for this transition \cite{obuse10}) relates the exponential decay of the mean conductance along the strip to dimensions of certain boundary operators. We extract these dimensions and compare them with the predictions of Ref.~\cite{bondesan11}.

\begin{figure}[t]
\centering
\includegraphics[]{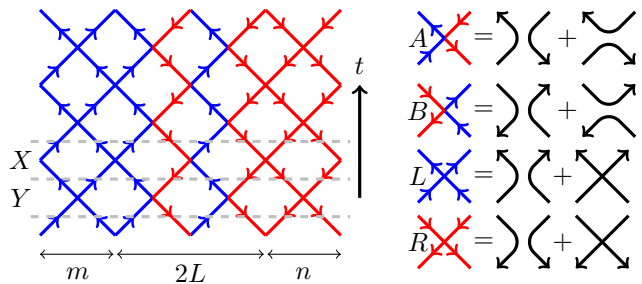}
\vspace{-10pt}
\caption{Left: Network model with bulk region of width $2L=4$, and $m=n=2$ extra channels at the left and right boundaries. Right: Schematic representation of the decomposition of average node transfer matrices. The two top rows show bulk nodes, see \eqref{eq:node-decomposition-bulk}, while the two bottom ones are boundary nodes, see \eqref{eq:node-decomposition-boundary}.}
\vspace{-15pt}
\label{fig:network}
\end{figure}

The network model for the SQH effect with extra edge channels is shown in Fig.~\ref{fig:network}. The bulk of width $2L$ contains alternating up- and down-going columns of links. In addition, $m$ ($n$) extra columns with the {\em same} chirality are added at the left (right) edge. These extra links can be directed up or down at either edge. We label the four possible variants by $({\cal L} = \pm m, {\cal R} = \pm n)$; positive labels $({\cal L}, {\cal R})$ mean the same direction of the edge links and the closest bulk link. The links of the network carry doublets of complex fluxes (labeled $\uparrow, \downarrow$) whose scattering on links is described by matrices uniformly distributed over the SU(2) group. The scattering at the two types of bulk nodes (labeled $S=A$ or $B$) is described by orthogonal matrices diagonal in spin indices:
\begin{equation}
{\cal S}_{S\up} = {\cal S}_{S\down} = \left( \begin{array}{cc} (1-t_S^2)^{1/2} & t_S^{\vphantom{2}} \\
-t_S^{\vphantom{2}} &  (1-t_S^2)^{1/2} \end{array} \right),
\label{eq:bulk-nodes}
\end{equation}
with $t_S$ the strength of the quantum tunneling. The SQH transition occurs when $t_A = t_B$.  The {\it boundary} nodes where the fluxes on the extra edge links scatter, are described by matrices of the form~\eqref{eq:bulk-nodes} but with two independent transmission amplitudes, $t_L$ and $t_R$, one for each edge of the system.

We employ the supersymmetry (SUSY) method for network models \cite{gruzberg99, read91, gruzberg97}, with modifications due to the extra edge links. The row-to-row transfer matrices $X$ and $Y$ (formed by multiplying all node transfer matrices, $T_A$ or $T_B$ for bulk nodes, $T_L$ and $T_R$ for boundary nodes, in a given row) act in the tensor product of bosonic and fermionic Fock spaces defined for each column of links. The columns form sites (labeled $i$) of a one-dimensional quantum system whose evolution in the vertical (imaginary time $t$) direction is given by the operator $U = \prod_t^{L_T} (XY)$, where $L_T$ is the number of $A$ nodes (or $B$ nodes) along the $t$-direction (see Fig.~\ref{fig:network}). With periodic boundary conditions in the $t$-direction, physical quantities, including conductance, may be written as correlation functions, $\langle \ldots \rangle \equiv{\rm STr} \big[ \,\ldots U \big]$, and the system is invariant under a global sl$(2|1)$ SUSY \cite{gruzberg99}.

Averaging over disorder independently on each link (we denote such averages by overbars) projects the Fock space of bosons and fermions onto the fundamental $V$ (dual-fundamental $V^\star$) 3-dimensional irreducible representation (irrep) of sl$(2|1)$ on up links (down links) \cite{gruzberg99}. The average node transfer matrices $\overline{T}_{S,i}$ act in the tensor products of {\it superspins} $V_i \otimes V^\star_{i+1}$ in the bulk and $V_i \otimes V_{i+1}$ or $V^\star_i \otimes V^\star_{i+1}$ in the extra edge regions. Thus, we have four types of supersymmetric {\it spin chains}
\begin{align}
  \label{eq:mon_charges_bd_1}
  (m;n) &:\quad V^{\otimes m} \otimes (V\otimes V^\star)^{\otimes L} \otimes \ (V^\star)^{\otimes n}, \\
  \label{eq:mon_charges_bd_2}
  (m;-n) &:\quad V^{\otimes m} \otimes (V\otimes V^\star)^{\otimes L} \otimes \ V^{\otimes n}, \\
  \label{eq:mon_charges_bd_3}
  (-m;n) &:\quad (V^\star)^{\otimes m} \otimes (V\otimes V^\star)^{\otimes L} \otimes (V^\star)^{\otimes n}, \\
  \label{eq:mon_charges_bd_4}
  (-m;-n) &:\quad (V^\star)^{\otimes m} \otimes (V\otimes V^\star)^{\otimes L} \otimes \ V^{\otimes n}.
\end{align}
All tensor products between neighboring sites of these chains decompose into two sl$(2|1)$ irreps. The averages $\overline{T}_{A,B}$ in the bulk and $\overline{T}_{L,R}$ at the boundaries read
\begin{align}
\label{eq:node-decomposition-bulk}
\overline{T}_{A,i} &= (1-t_A^2) I + t_A^2 E_i, && A \leftrightarrow B, \\
\overline{T}_{L,i} &= (1-t_L^2) I + t_L^2 P_{i,i+1}, && L \leftrightarrow R,
\label{eq:node-decomposition-boundary}
\end{align}
where $I$ is the identity, $E_i$ projects onto the singlet in the decomposition of $V_i \otimes V^\star_{i+1}$ (and $V^\star_i \otimes V_{i+1}$), and $P_{i,i+1}$ is the graded permutation of states on sites $i$ and $i+1$. The chain \eqref{eq:mon_charges_bd_1} with $m=n$ corresponds to a sigma-model with topological angle $\Theta=(2n+1)\pi$ and to a transition between the $n$-th and the $(n+1)$-st plateau.

The decompositions (\ref{eq:node-decomposition-bulk}, \ref{eq:node-decomposition-boundary}) have a natural graphical representation shown schematically (without coefficients) in Fig.~\ref{fig:network} (right). At a bulk node links can be separated into disjoint lines in two ways, while at a boundary node the lines can avoid each other or cross. Multiplying the transfer matrices to calculate the partition function, the result is the sum of all contributions of dense closed loops of weight one filling the links of the network, weighted by factors of either $t_S^2$ or $(1-t_S^2)$ for each node. In the bulk the loops are percolation hulls, and the loop model is equivalent to bond percolation on a square lattice \cite{gruzberg99}. The presence of extra edges generalizes this non-trivially, since loops can intersect at the boundary. In this situation all configurations of the loop model fall into disjoint sectors \cite{martin1991potts} labeled by the number $\ell + 2k$ of {\it through lines} extending throughout the system in the $t$-direction, where $\ell = |{\cal L} - {\cal R}|$ is the minimal possible number of the through lines for given ${\cal L}, {\cal R}$.

In the strip geometry our model has external links at the top and bottom, where we can insert or extract current. In this geometry the mean conductance ${\bar g}^{{\cal L}, {\cal R}}$ between the top and bottom contacts is the average number of through lines going from the source to the drain of the current, times $2$ for the spin. With the source at the bottom, the minimal number of available through lines is $k^{{\cal L}, {\cal R}}_{\text{min}} = \max(0,{\cal L} - {\cal R})$, whence
\begin{align}
\label{eq:g-bar}
\bar{g}^{{\cal L}, {\cal R}} = 2 k^{{\cal L}, {\cal R}}_{\text{min}} + 2 \sum\nolimits_{k=1}^\infty k P(k,L_T/L, L_T/\xi),
\end{align}
where $P(k,L_T/L,L_T/\xi)$ is the probability (symmetric in ${\cal L}$ and ${\cal R}$) that exactly $2k$ ``paired'' through lines run through the system of size $2L_T$ by $2L+m+n$, and $\xi$ is the bulk correlation length. At the transition, $\xi=\infty$, and for large $L_T/L$ we expect that $P(k,L_T/L,0) \sim e^{-\pi h^{{\cal L}, {\cal R}}(k) L_T/L}$. Conformal invariance at the transition allows to identify the exponents $h^{{\cal L}, {\cal R}}(k)$ with dimensions of certain boundary operators. In the literature on self-avoiding walks such exponents are called watermelon exponents, and the through lines are called legs. For $L_T/L \gg 1$ the sum in Eq.~\eqref{eq:g-bar} for $\bar{g}$ is dominated by $k=1$, and we denote $\galr$ the exponent of the first subleading correction:
\begin{align}
\bar{g}^{{\cal L}, {\cal R}} \sim 2 k^{{\cal L}, {\cal R}}_{\text{min}}
+ C_1 e^{-\pi h^{{\cal  L}, {\cal R}}(1) \frac{L_T}{L}}
+ C_2 e^{-\pi \galr \frac{L_T}{L}}.
\label{eq:g-bar_asymptotic}
\end{align}

Critical properties of the above geometric loop model, including the exponents $h^{{\cal L}, {\cal R}}(k)$, can be extracted from the anisotropic limit of the spin chains (\ref{eq:mon_charges_bd_1}--\ref{eq:mon_charges_bd_4}), obtained by taking all $t_S \ll 1$. In this limit the evolution operator in one unit of time becomes $\overline{XY} \approx \exp\big(-t_A t_B H\big)$. For definiteness we focus on the case~\eqref{eq:mon_charges_bd_1}; the critical ($t_A = t_B$) Hamiltonian $H$ is then
\begin{equation}
\label{eq:hamiltonian}
H = - u \sum_{i=0}^{m-1} P_{i,i+1} - \!\!\! \sum_{i=m}^{2L+m-2} \!\!\! E_i - \,\,
v \!\!\!\! \sum_{i=2L+m-1}^{2L+m+n-2} \!\!\!\! P_{i,i+1}.
\end{equation}
The interaction between the superspins in the bulk is antiferromagnetic and uniform at the critical point. Interactions between boundary spins are ferromagnetic (compare with \cite{gruzberg97}), and their magnitudes $u = (t_L/t_A)^2$, $v = (t_R/t_A)^2$ are kept as arbitrary {\it positive} numbers. From the diagonalization of $H$, described in \cite{bondesan11} along with full details of the derivation of the results presented below, we obtain the critical exponents $h^{m,n}(k)$ as scaling limits of the lowest eigenvalues in a given sector. A more detailed knowledge of $\bar{g}^{{\cal L},{\cal R}}$ is needed for predicting $\galr$ in Eq.~\eqref{eq:g-bar_asymptotic} analytically \cite{subleading}, but in most cases we can identify its numerical value as $h^{{\cal L},{\cal R}}(2)$.

The exact expressions for $h^{m,n}(k)$ derived in \cite{bondesan11} appear in Table \ref{tab:crit_exp}. There $h_{r,s} = \big[(3r - 2s)^2 - 1\big]/24$, $\minlr=\min(m,n)$ and the parameter $r_k$ is given in terms of $m, n$ and $k$ for $m\ge n$ (if $m<n$ exchange $m$ and $n$) by
\begin{equation}
\label{eq:ri}
r_k = \frac{6}{\pi} \arccos \left(\frac{\sqrt{3}}{2} \sqrt{ \frac{(m+1+k)(n+1-k)}{(m+1)(n+1)}} \right).
\end{equation}

A few remarks are in order. The exponents are independent of the boundary couplings $u, v$, implying a boundary renormalization group flow to the stable fixed point with infinite boundary couplings, similar to Ref.~\cite{boundaryRG}. Moreover, since $u,v$ are positive (ferromagnetic), randomness in $u,v$ does not change the exponents. If the number of legs is $\ell$ or $m+n+2j$ with $j > 0$, the exponents equal those of critical percolation with respectively $0$ or $2j$ hulls through the system (in particular, this is so when $m=0$ or $n=0$). However, when the number of legs is between $\ell+2$ and $m+n$, exponents are highly non-trivial and, remarkably, they are irrational.
\begin{table}[t]
  \centering
  \begin{tabular}{|c|c|c|}
    \hline
    $k$ & $\#(\mbox{legs}) = \ell + 2k$ & $h^{m,n}(k)$ \\
    \hline
    $0$ & $\ell$ & $h_{r_0,r_0} = 0$ \\
    $1$ & $\ell+2$ & $h_{r_1,r_1}$ \\
    $\vdots$ & $\vdots$ & $\vdots$ \\
    $\minlr$ & $n+m$ & $h_{r_{\minlr},r_{\minlr}}$ \\
    $\minlr+1$ & $n+m+2$ & $h_{1,3} = 1/3$ \\
    $\vdots$ & $\vdots$ & $\vdots$ \\
    $\minlr+j$ & $n+m+2j$ & $h_{1,1+2j} = j(2j-1)/3$ \\
    $\vdots$ & $\vdots$ & $\vdots$ \\
    \hline
  \end{tabular}
\caption{The watermelon exponents. $k$ is the number of paired through lines on top of $\ell$ unpaired ones and $\minlr=\min(m,n)$.   $h_{r,s}$ is the Kac table and $r_k$ is defined as in Eq.~\eqref{eq:ri}.}
\vspace{-15pt}
\label{tab:crit_exp}
\end{table}

The exponents $h^{{\cal L}, {\cal R}}(k)$ are symmetric in ${\cal L}$ and ${\cal R}$ by invariance of the spectrum under left-right reflection. Further exponent relations follow from symmetries of the critical spin chains (\ref{eq:mon_charges_bd_1}--\ref{eq:mon_charges_bd_4}). In each case $V \otimes V^\star$ (and $V^\star \otimes V$) interacts through $E_i$, and $V \otimes V$ (and $V^\star \otimes V^\star$) through $P_i$, and the $u, v$ can be set to one thanks to the universality. Then the top-bottom reflection switching $V$ and $V^\star$ induces a mapping between the chains (sometimes with different lengths of the bulk region):
$({\cal L}, {\cal R}) \leftrightarrow (-{\cal L}-1, -{\cal R}-1)$. This implies:
\begin{align}
h^{{\cal L}, {\cal R}}(k) = h^{-{\cal L}-1, -{\cal R}-1}(k).
\label{eq:symmetry}
\end{align}
In particular, this relates pairwise exponents for the chains (\ref{eq:mon_charges_bd_1}, \ref{eq:mon_charges_bd_4}), as well as for (\ref{eq:mon_charges_bd_2}, \ref{eq:mon_charges_bd_3}): $h^{-m,-n}(k) = h^{m-1,n-1}(k)$, $h^{-m,n}(k) = h^{m-1,-n-1}(k)$. When the total number of legs in the model (\ref{eq:mon_charges_bd_4}) equals $m + n + 2j$ with $j \geqslant 0$, the corresponding exponent $h^{m-1,n-1}(k) = h_{1,3+2j} = (j+1)(2j+1)/3$, since in this case we must write the total number of legs as $(m-1) + (n-1) + 2j + 2$.

Finally, consider the chain \eqref{eq:mon_charges_bd_2}. Regarding the leftmost site on the right boundary as part of the bulk, and noting that we have identical chiralities at the two boundaries (so that $\ell = m+n$), the $(m+n+2j)$-leg exponents are simply $h_{1,2+2j} = j(2j+1)/3$, $j=0,1,\dots$ independently of $m$ and $n$.  Results for \eqref{eq:mon_charges_bd_3} easily follow from \eqref{eq:symmetry}: the $(m+n+2j)$-leg exponents are $h_{1,2+2j} = j(2j+1)/3$ in both models (\ref{eq:mon_charges_bd_2}, \ref{eq:mon_charges_bd_3}).

\begin{table}[t]
\begin{tabular}{|c| c c l l| r r|}
\hline
\multirow{2}{*}{${\cal L},{\cal R}$} &
\multicolumn{4}{c|}{numerical simulations} &
\multicolumn{2}{c|}{analytical} \\
\cline{2-7}
  & $2L$ & $\theta_e$ & $h^{{\cal  L}, {\cal R}}(1)$ &$\galr$
  & $h^{{\cal  L}, {\cal R}}(1)$ &$h^{{\cal  L}, {\cal R}}(2)$ \\
\hline\hline
$0,0$ &$\infty$ & -  &$0.3333(12)$& $2.06(63)$ & $1/3$  & $2$ \\
\hline
\hline
$0,1$ &$\infty$ & $\theta_c$ &  $0.3330(7)$ & $2.00(36)$ &$1/3$ &$2$\\
$0,1$ &$\infty$ & $\pi/10$   & $0.3341(31)$& $2.09(47)$ & $1/3$ &$2$\\
$0,1$ &$\infty$ & rand   & $0.3332(10)$ & $1.99(55)$& $1/3$ &$2$\\
\hline
$0,10$ &$\infty$ & $\theta_c$ & $0.3325(24)$& $1.81(55)$  & $1/3$ &$2$ \\
$0,10$ &$\infty$ & $\pi/10$  & $0.3318(24)$& $2.03(43)$ & $1/3$ &$2$\\
\hline\hline
$1,1$ & $\infty$ & $\theta_c$  & $0.03775(25)$& $0.333(5)$& $0.037720$&$1/3$ \\
$1,1$ & $\infty$ & $\pi/10$  & $0.03779(31)$&$0.319(13)$ & $0.037720$&$1/3$\\
$1,1$ & $\infty$ & rand  & $0.03773(45)$&$0.343(10)$ & $0.037720$&$1/3$\\
\hline
$2,2$ &$50$ & $\theta_c$ & $0.01600(2)$& $0.0737(6)$  &$0.015906$& $0.0732$\\
\hline
$3,3$ &$50$ & $\theta_c$ & $0.00880(1)$& $0.0396(2)$ &$0.008797$ &
 $0.0377$\\
\hline
$4,4$ &$50$ & $\theta_c$ & $0.00562(1)$& $0.0255(1)$ &$0.005587$ & $0.0233$\\
\hline
$1,2$ &$\infty$ & $\theta_c$ & $0.0520(25)$&$0.334(43)$ &$0.052083$ & $1/3$\\
$1,2$ &$\infty$ & $\pi/10$  & $0.0517(10)$ &$0.307(16)$ &$0.052083$ & $1/3$\\
$1,2$ &$50$ & rand \ \   & $0.05221(3)$&$0.338(8)$ &$0.052083$ & $1/3$\\
\hline
$1,3$ &$50$ & $\theta_c$  & $0.05986(3)$& $0.339(9)$ &$0.059697$&$1/3$ \\
\hline
$1,4$ &$50$ & $\theta_c$ & $0.06425(3)$& $0.34(1)$&$0.064421$& $1/3$\\
\hline
$2,3$ &$50$ & $\theta_c$  & $0.02449(10)$& $0.088(1)$ &$0.024348$& $0.0847$\\
$2,3$ &$\infty$ & $\pi/10$  &$0.02476(42)$ & $0.086(3)$ &$0.024348$& $0.0847$\\
\hline
$2,4$ &$50$ & $\theta_c$  & $0.02954(7)$& $0.095(1)$&$0.029589$ & $0.0920$\\
\hline
\hline
$-2,-2$ & $\infty$ & $\pi/3$ & $0.0377(4)$ & $0.339(10)$ & $0.037720$
 & $1/3$\\
$-2,-2$ & $\infty$ & $\pi/10$ & $0.0372(8)$ & $0.281(4)$ & $0.037720$
 & $1/3$ \\
\hline
$-3,-3$ & $\infty$ & $\pi/3$ & $0.01611(9)$ & $0.0749(7)$ & $0.015906$
 & $0.0732$\\
$-3,-3$ & $\infty$ & $\pi/10$ & $0.0160(9)$  & $0.0730(30)$  & $0.015906$
 & $0.0732$ \\
\hline
$-3,-2$ & $\infty$ & $\pi/3$ & $0.0522(2)$ & $0.333(13)$ & $0.052083$ &
 $1/3$ \\
$-3,-2$ & $\infty$ & $\pi/10$ &$0.0549(30)$ & $0.336(13)$ & $0.052083$ &
 $1/3$ \\
\hline\hline
$-1,0$ &$\infty$ & $\theta_c$ & $0.999(9)$ & --- 
&$1$& $10/3$
\\
\hline
$-2,0$ &$\infty$ & $\theta_c$ & $0.999(3)$ & --- 
&$1$ & $10/3$ \\
\hline
$-2,1$ &$\infty$ & $\theta_c$ & $0.998(3)$ & --- 
&$1$ & $10/3$ \\
\hline
$-2,2$ &$\infty$ & $\theta_c$ & $0.993(1)$ & --- 
&$1$ & $10/3$ \\
\hline
\end{tabular}
\caption{
Numerically obtained exponents $h^{{\cal  L}, {\cal R}}(1), \gamma^{\LMon,\RMon}$ and analytical predictions for $h^{\LMon,\RMon}(1), h^{\LMon,\RMon}(2)$ for various values of $({\cal L},{\cal R})$. $\infty$ means that the numerical exponents are obtained by finite size scaling \cite{supplement}. ``rand'' indicates randomly distributed $\theta_e$. ``---'' means that numerical estimates of these subleading exponents are unreliable.
}
\label{tab:numerics}
\vspace{-15pt}
\end{table}

We now present extensive numerical simulations to verify our analytical predictions for the exponents $h^{{\cal L}, {\cal R}}(1)$, the independence on $t_{L,R}$, the symmetry relation \eqref{eq:symmetry}, and to determine the subleading exponent $\gamma^{\LMon,\RMon}$. We numerically calculate the conductance of critical SQH networks with extra channels in the strip geometry with length $2L_T$ and width $2L+m+n$. We parametrize the transmission amplitude in Eq.~\eqref{eq:bulk-nodes} as $t_{A}=\sin \theta$ and $t_B=\cos \theta$. In terms of $\theta$ the SQH transition occurs at $\theta = \theta_c \equiv \pi/4$. Similarly, we write $t_{L,R}=\sin \theta_e$, where $\theta_e$ can be arbitrarily tuned or chosen randomly in $[0,\pi]$ independently for each boundary node.

The transmission matrix $\bm{t}$ for the SQH network model in the strip geometry is effectively calculated using the transfer matrix method \cite{kramer05}; the conductance is given by the Landauer formula: $g={\rm Tr}\ \bm{t} \bm{t}^\dagger$. Ideally the exponents $h^{{\cal L}, {\cal R}}(1)$ and $\galr$ would be obtained by fitting the data to Eq.~\eqref{eq:g-bar_asymptotic}. However, finite-size effects hamper such analysis in actual simulations unless $\theta_e=\theta_c$ or  $\theta_e$ is random. Therefore, we apply finite size-scaling analysis to systems of various widths \cite{supplement}. We only simulate networks corresponding to chains (\ref{eq:mon_charges_bd_1}, \ref{eq:mon_charges_bd_3}, \ref{eq:mon_charges_bd_4}) with ${\cal L}\le {\cal R}$, so that $k^{{\cal L}, {\cal R}}_{\text{min}}=0$. Table\ \ref{tab:numerics} summarizes our numerical results, and includes  analytical predictions for comparison.

First we focus on systems~(\ref{eq:mon_charges_bd_1}) with $\LMon=0, \RMon\ge0$.  As shown in Table \ref{tab:numerics}, numerically obtained $h^{0,\RMon}(1)$ agrees with analytical results and $\gamma^{0,\RMon}$ coincides with $h^{0,\RMon}(2)$ for various $\RMon$ and $\theta_e$.  This suggests that the level-one descendant of $h^{0,\RMon}(1)$, $h^{0,\RMon}(1)$+1=4/3, does not contribute to the conductance in these systems.  In addition, $h^{0,1}(1) = 1/3$ and $\gamma^{0,1}=2$ are verified even for random $\theta_e$, confirming our expectation that randomness in the boundary couplings is irrelevant.

Next, we consider cases $\LMon, \RMon > 0$, where $h^{\LMon,\RMon}(1)$ is irrational.  Since the $L$ dependence is weak for $\theta_e=\theta_c$ and random $\theta_e$, we extract several $h^{m,n}(k)$ for these $\theta_e$ without finite-size scaling analysis. We confirm that the numerically obtained exponents $h^{m,n}(1)$ and $\gamma^{m,n}$ agree well with the analytical $h^{m,n}(1)$ and $h^{m,n}(2)$, respectively.  Note that there are no analytically predicted exponents with values between $h^{m,n}(1)$ and $h^{m,n}(2)$.

Furthermore, we study the cases $\LMon, \RMon < 0$, whose exponents $h^{\LMon,\RMon}(k)$ are related to those of systems with $\LMon, \RMon > 0$ by Eq.~(\ref{eq:symmetry}). Comparing results for $(-m,-n)$ and $(n-1,m-1)$ in Table \ref{tab:numerics} (and assuming the $\LMon \leftrightarrow \RMon$ symmetry), we see that Eq.~(\ref{eq:symmetry}) for $k=1,2$ is verified even if $\theta_e \ne \theta_c$. Finally, when $\LMon < 0$, $\RMon \ge 0$ in Eq.~(\ref{eq:mon_charges_bd_3}) our numerics confirm $h^{\LMon,\RMon}(1)=1$ for various cases. Results for the subleading exponent $\galr$ are presently inconclusive because of large numerical errors.

Our comprehensive numerical simulations confirm the exact analytical predictions for the leading  exponent $h^{{\cal L}, {\cal R}}(1)$ and, in most cases, allow us to identify the first subleading exponent $\gamma^{\LMon,\RMon}$ as $h^{{\cal L}, {\cal R}}(2)$.

In conclusion, we have considered the SQH transition with extra edge channels. We have mapped the corresponding network model to a classical loop model, whose boundary critical exponents have recently been obtained exactly. Using the mapping we obtain exact critical exponents at the SQH transition from the exponential decay of the mean conductance in the strip geometry. Our extensive numerical simulations confirm the analytical results for the boundary exponents. The demonstrated influence of extra edge channels on boundary critical behavior should be broadly applicable to other QH transitions.

\paragraph{Acknowledgments.}
The numerical simulations have been performed using the PADS resource (NSF grant OCI-0821678) and the Teraport Cluster at the Computational Institute at the University of Chicago.  The work of J.~L.~J.  and H.~S.  was supported by a grant from the ANR Projet 2010 Blanc SIMI 4: DIME.  H.~O. is supported by Grant-in-Aid for JSPS for Young Scientists. I. A. G. was supported by NSF Grants No. DMR-0448820 and No. DMR-0213745. H.~S. thanks C. Candu, N. Read and V. Schomerus for discussions. H.~O. thanks F. Evers for discussions.


\clearpage
\newpage
 \setcounter{equation}{0}
 \setcounter{table}{0}
 \setcounter{figure}{0}
\setcounter{page}{1}
\thispagestyle{empty}

\section{Supplemental material for
``Exact exponents for the spin quantum Hall transition in the presence of multiple edge channels''}

In this supplemental material, we explain the details of the numerical
finite-size scaling analysis that we use to calculate the leading
and the first subleading exponents, $\honelr$ and $\galr$,
respectively, from the conductance of the spin quantum Hall network
model with multiple edge channels in the strip geometry.

\subsection{Conductance in the strip geometry}

The ensemble averaged conductance $\bar{g}^{\LMon,\RMon}$ is predicted (see Eq.\ (9) in the main paper) to decay as
\begin{align}
\bar{g}^{{\cal L}, {\cal R}} \sim 2 k^{{\cal L}, {\cal R}}_{\text{min}}+  C_1 e^{-\pi h^{{\cal  L}, {\cal R}}(1)  L_T/L}
+ C_2 e^{-\pi \galr  L_T/L},
\label{eq:g-bar_asymptotic_supp}
\end{align}
where $2L_T$ and $2L$ denote the system's length and the bulk width,
respectively. Therefore, the exponents $\honelr$ and $\galr$ are
numerically calculated by fitting the averaged conductance
$\bar{g}^{\LMon,\RMon}$ to Eq.\ (\ref{eq:g-bar_asymptotic_supp}) as a
function of the aspect ratio $L_T/L$.  In the actual numerical
simulations, we typically vary $L_T/L$ in the range $[2,40]$ for systems
for which the predicted exponent $h^{\LMon,\RMon}(1)< 1/3$, and the
number of disorder realizations is then taken to be ${\cal O}(10^5) \sim
{\cal O}(10^6)$. On the other hand, since the conductance rapidly decays
for systems with the predicted value $h^{\LMon,\RMon}(1) \ge 1/3$, the
range of $L_T/L$ for such systems is varied between 1 and 4, and the
number of samples is taken to be ${\cal O}(10^6) \sim {\cal O}(10^7)$.
As we have mentioned in the main paper, in our numerical study we only
consider the cases of systems for which the minimal number of available
through lines is $k^{{\cal L}, {\cal R}}_{\text{min}} = 0$.

\begin{figure*}[t]
\includegraphics[width=0.9\textwidth]{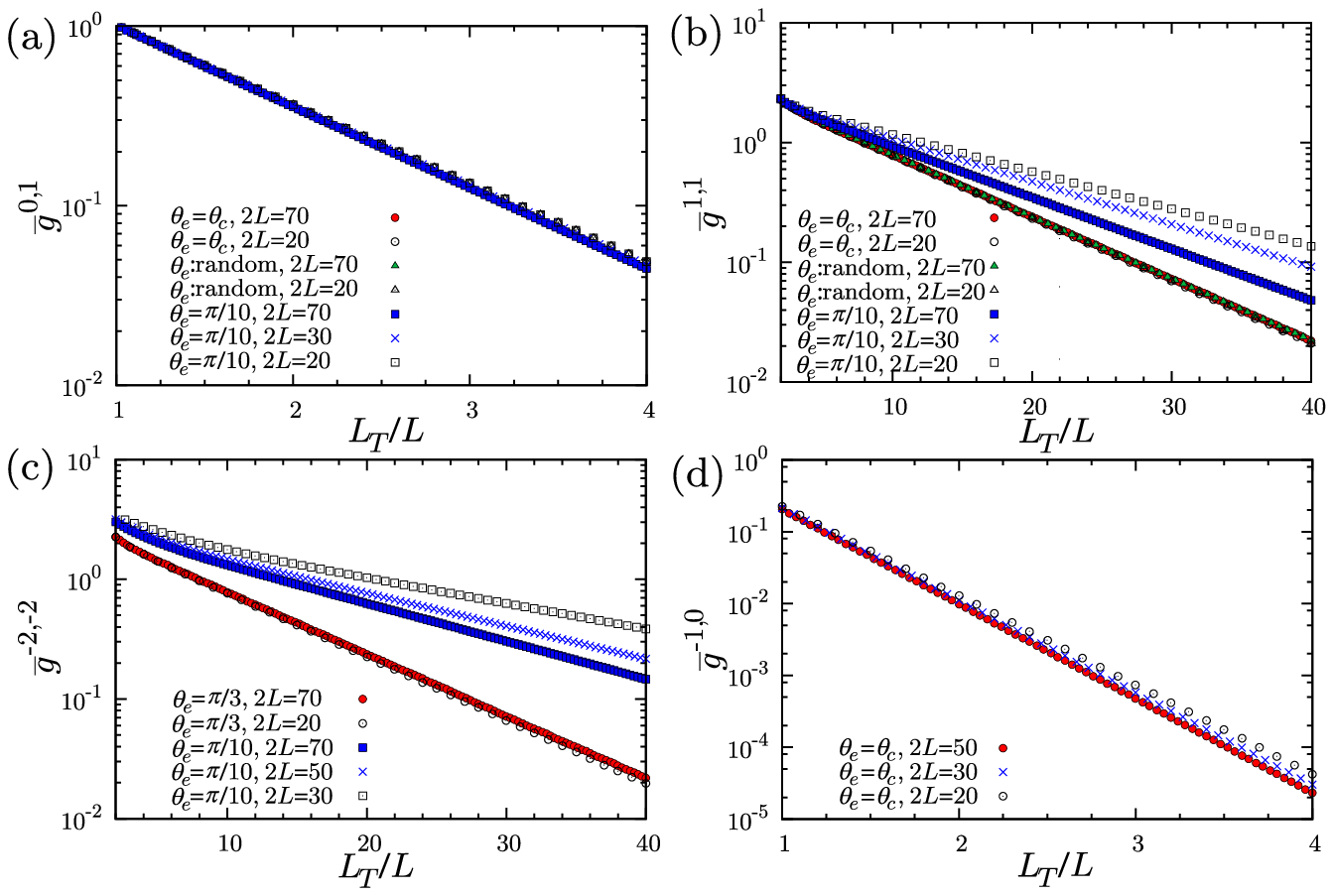} 
\caption{ The averaged conductance $\bar{g}^{\LMon,\RMon}$ as a function of $L_T/L$ for systems with different width $2L$ and $\theta_e$, as indicated in the figure, in the cases of (a) $(\LMon,\RMon)=(0,1)$, (b) $(1,1)$, (c) $(-2,-2)$, and (d) $(-1,0)$. Note that the error bars are smaller than the symbols.}
\label{fig:conductance}
\end{figure*}

Figure \ref{fig:conductance} shows the $L_T/L$ dependence of the
averaged conductance $\bar{g}^{\LMon,\RMon}$ for several systems with
different sets of $\LMon$ and $\RMon$ as well as different values of the
width and $\theta_e$. Although Eq.\ (\ref{eq:g-bar_asymptotic_supp}) predicts
that the conductance decays exponentially with the universal exponent
$\honelr$ in the long strip geometry $(L_T \gg L)$, the averaged
conductance calculated numerically exhibits an exponential decay with
exponents that depend on the width and $\theta_e$, except for the cases
of $\LMon=0, \RMon=1$ and $\LMon=\RMon=1$ with $\theta_e=\theta_c$ or
randomly distributed $\theta_e$.

\begin{table}[h]
\caption{
A list for the exponents $\honelrL$ and $\galrL$ calculated by fitting the conductance of the finite width $L$ to Eq.\ (\ref{eq:g-bar_asymptotic_supp}) for various $(\LMon,  \RMon)$ and $\theta_e$.
\label{tab:exponent}
}
\begin{tabular}{c c |l l|l  l}
\hline\hline
$({\cal L},{\cal R})$&
$2L$&
$h_L^{{\cal L},{\cal R}}(1)$ & $\gamma_L^{{\cal L},{\cal R}}$ &
$h_L^{{\cal L},{\cal R}}(1)$ & $\gamma_L^{{\cal L},{\cal R}}$ \\
\hline
&&
\multicolumn{2}{c|}{$\theta_e=\theta_c$}&
\multicolumn{2}{c}{$\theta_e=\pi/10$}\\
\cline{3-6}
$(0,1)$ & $70$ &  $0.3300(1)$ & $2.03(11)$ & $0.3301(4)$ & $2.03(16)$\\
 & $50$ &  $0.3284(2)$ & $1.90(12)$ & $0.3289(5)$ & $1.96(41)$\\
 & $30$ &  $0.3254(7)$ & $1.90(47)$ & $0.3260(6)$ & $2.04(53)$\\
\hline
$(0,10)$ & $60$ &  $0.3288(3)$ & $1.67(24)$ & $0.3289(3)$ & $2.08(42)$\\
 & $50$ &  $0.3285(3)$ & $1.98(33)$ & $0.3282(4)$ & $1.65(34)$\\
 & $30$ &  $0.3249(3)$ & $1.79(30)$ & $0.3252(4)$ & $1.78(60)$\\
\hline
$(1,1)$ & $70$ &  $0.03789(1)$ & $0.333(2)$ & $0.03147(3)$ & $0.338(5)$\\
 & $50$ &  $0.03794(1)$ & $0.329(3)$ & $0.02971(4)$ & $0.339(7)$\\
 & $30$ &  $0.03811(2)$ & $0.330(3)$ & $0.02610(2)$ & $0.334(3)$\\
\hline
$(1,2)$ & $70$ &  $0.05210(3)$ & $0.333(7)$ & $0.04007(2)$ & $0.344(6)$\\
 & $50$ &  $0.05248(3)$ & $0.335(8)$ & $0.03697(2)$ & $0.347(4)$\\
 & $30$ &  $0.05266(4)$ & $0.332(9)$ & $0.03144(2)$ & $0.357(4)$\\
\hline
$(-1,0)$ & $50$ &  $0.9637(5)$ & $2.0(5)$ &  &\\
 & $30$ &  $0.9400(4)$ & $2.2(7)$ &  & \\
 & $20$ &  $0.9109(4)$ & $2.4(9)$ &  & \\
\hline
\hline
&&
\multicolumn{2}{c|}{$\theta_e=\pi/3$}&
\multicolumn{2}{c}{$\theta_e=\pi/10$}\\
\cline{3-6}
$(-2,-2)$ & $70$ &  $0.03796(3)$ & $0.326(5)$ & $0.02315(5)$ & $0.149(1)$\\
 & $50$ &  $0.03819(4)$ & $0.336(6)$ & $0.02010(6)$ & $0.123(1)$\\
 & $30$ &  $0.03855(5)$ & $0.335(7)$ & $0.01553(14)$ & $0.087(1)$\\
\hline
$(-3,-3)$ & $70$ & $0.01605(2)$&$0.0749(4)$&  $0.00968(3)$ & $0.0460(4)$\\
 & $50$ &$0.01605(3)$ & $0.0748(6)$ & $0.00865(5)$ & $0.0422(5)$ \\
 & $30$ & $0.01608(7)$ & $0.0760(30)$ & $0.00685(7)$ & $0.0344(6)$ \\
\hline\hline
\end{tabular}
\end{table}

\begin{figure}[h]
\includegraphics[width=8.5cm]{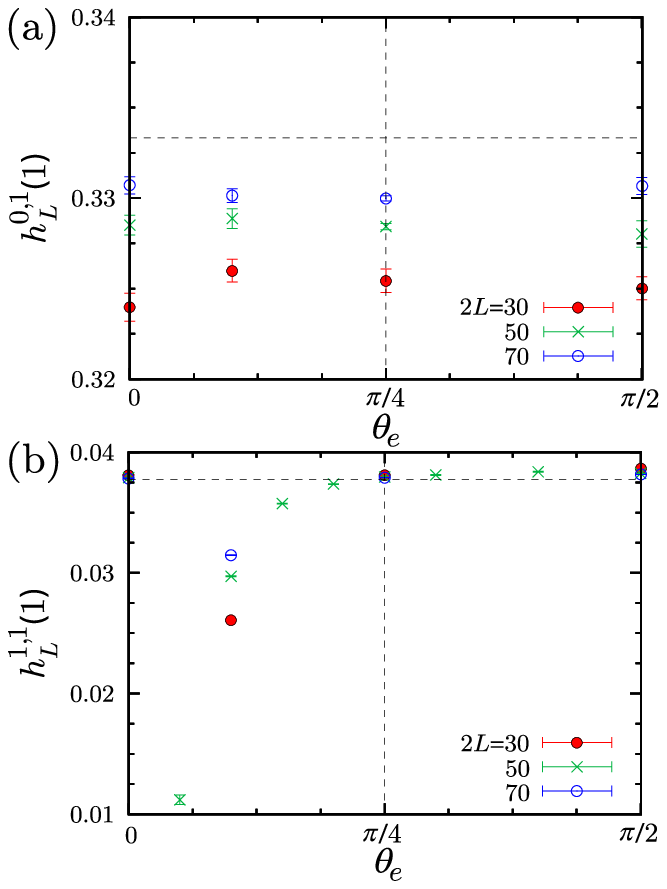}
\caption{The $\theta_e$ dependence of (a) $h_L^{0,1}(1)$ and (b) $h_L^{1,1}(1)$ for $2L=30,50,70$.  The horizontal dashed lines represent the analytically predicted values of (a) $h^{0,1}(1)=1/3$ and (b)  $h^{1,1}(1)\approx0.037720$. The vertical dashed line indicates $\theta_c$. The symbols at $\theta_e=0$ represent $h_L^{0,1}(1)$ and $h_L^{1,1}(1)$ for the system with randomly distributed $\theta_e$.}
\label{fig:theta_dep_hmn}
\end{figure}

To determine the width dependence of the exponents, we calculate
$\honelr$ and $\galr$ by fitting the conductance for each width to Eq.\
(\ref{eq:g-bar_asymptotic_supp}) for systems with various sets of
$(\LMon,\RMon)$ and several $\theta_e$. Hereafter, in order to avoid
confusion, we introduce the notations $h_L^{{\cal L}, {\cal R}}(1)$ and
$\gamma_L^{{\cal L}, {\cal R}}$ for the exponents determined in this
way, to stress the effects of the finite width $L$ on $\honelr$ and
$\galr$, respectively.  We also introduce a fitting parameter $k^{{\cal
L}, {\cal R}}_L$ replacing the theoretical value $k^{{\cal L}, {\cal
R}}_{\text{min}}=0$ of the conductance of an infinitely long strip.  In
each case we ensure the goodness of fit to be almost one and confirm
that $k^{{\cal L}, {\cal R}}_L$ is almost zero.  The fitting results are
summarized in Table \ref{tab:exponent}. Figure \ref{fig:theta_dep_hmn}
highlights the width and $\theta_e$ dependence of the exponents
$h_L^{0,1}(1)$ and $h_L^{1,1}(1)$. As shown in Fig.\
\ref{fig:theta_dep_hmn} (a), even in the case of ${\cal L}=0$ and ${\cal
R}=1$, $h_L^{0,1}(1)$ depends slightly on the width for all tested
$\theta_e$, although the exponent seems to approach the analytical
prediction $h^{0,1}(1)=1/3$ with increasing width $L$.

In contrast, $h_L^{1,1}(1)$ is  more sensitive to the value of $\theta_e$ as shown in Fig.\ \ref{fig:theta_dep_hmn} (b). For $\theta_e < \theta_c$, $h_L^{1,1}(1)$ shows a strong width dependence upon approaching the analytically predicted $h^{1,1}(1)$. For $\theta_e \ge \theta_c$, $h_L^{1,1}(1)$ is close to the analytical prediction even for small widths. Remarkably, the width dependence is significantly diminished when $\theta_e$ is set as $\theta_c$ or randomly distributed at each node. This diminishing is confirmed for other systems with $\LMon,\RMon >0$. We note that $\galrL$ also shows the width and $\theta_e$ dependence in some cases, as found from Table \ref{tab:exponent}.

Summarizing this section, we have found that the exponents $\honelrL$ and $\galrL$ calculated from the conductance in finite size systems depend on the system width, and differ from the analytical predictions due to finite-size effects.

\subsection{Finite size scaling analysis}

Here we explain how we use a finite-size scaling analysis to calculate $\honelr$ and $\galr$ in the thermodynamic limit from the conductance of finite systems. The fact that $\honelrL$ and $\galrL$ depend on the width $L$ implies that there are irrelevant fields at the critical point of the SQH network model with extra edge channels. Here, we generalize Eq.\ (\ref{eq:g-bar_asymptotic_supp}) to include contributions of the leading irrelevant field as follows. 

\begin{enumerate}[i.]
\item  In the standard finite-size scaling analysis of the Anderson transition \cite{Slevin99}, a Lyapunov exponent describing the exponential growth or decay of eigenvalues of the transfer matrix in the long quasi-one dimensional geometry is considered as a quantity described by a scaling function. By analogy, we assume that $\honelr$ and $\galr$ in Eq.\ (\ref{eq:g-bar_asymptotic_supp}) are also described by scaling functions of the width $L$, such as
\begin{eqnarray}
h_L^{{\cal L}, {\cal R}}(1) &=& f_h\left[(2L)^{1/\nu} u_0(x),(2L)^{y}u_1(x)\right],\\
\gamma_L^{{\cal L},{\cal R}} &=& f_\gamma\left[(2L)^{1/\nu} u_0(x),(2L)^{y}u_1(x)\right],
\label{eq:scaling_function}
\end{eqnarray}
where $u_0(x)$ and $u_1(x)$ represent the relevant and irrelevant scaling fields, respectively, and $\nu$ and $y(<0)$ are their scaling exponents.
The variable $x$ represents the distance from the critical point. 

\item Since we focus only on the critical point ($x=0$), $u_0(0)=0$ and $u_1(0)$ becomes a constant. Then, the scaling functions $f_{h,\gamma} \left[0,(2L)^{y}u_1(0)\right]$ depend only on the irrelevant field $u_1$. We expand them in Taylor series up to the order $N^h$ ($N^{\gamma}$) as
\begin{eqnarray}
f_h\left[0,(2L)^{y}u_1(0)\right] &=& \sum_{k=0}^{N^h} a_k^h (2L)^{ky},\\ f_\gamma\left[0,(2L)^{y}u_1(0)\right] &=& \sum_{k=0}^{N^\gamma} a_k^\gamma (2L)^{ky}.  
\label{eq:irrelevant}
\end{eqnarray}
Note that $a_0^h = h_\infty^{{\cal L}, {\cal R}}(1)\equiv \honelr$ and $a_0^\gamma= \gamma_\infty^{{\cal L},{\cal R}} \equiv \galr$.

\item Since the prefactors $C_{1,2}$ in Eq.\ (\ref{eq:g-bar_asymptotic_supp}) also depend on the width $L$, we take account of this finite-size effect by considering
$C_{1,2}$ as functions of $1/2L$ and expanding them in Taylor series up to the order $N_C^h$ ($N_C^{\gamma}$). 
\end{enumerate}

According to all these assumptions we write down the scaling function for the average conductance as
\begin{align}
& \bar{g}^{\LMon,\RMon} = 2 k^{{\cal L}, {\cal R}}_L +
\sum_{l=0}^{N_{C}^{h}} F_l^{h}(L,L_T) + \sum_{l=0}^{N_{C}^{\gamma}} F_l^{\gamma}(L,L_T), \nonumber \\
& F_l^{\text{x}=h,\gamma}(L,L_T) \equiv \frac{C_l^\text{x}}{(2L)^l} \exp\left[-\pi\sum_{k=0}^{N^\text{x}} a_k^\text{x} (2L)^{ky} \frac{L_T}{L}\right],
\label{eq:scaling-function}
\end{align}
where $k^{\LMon,\RMon}_L, y, a_k^h, a_k^\gamma, C_l^h, C_l^\gamma$ are treated as fitting parameters.

In our actual calculations, we set $N^h=N^\gamma=2$ and $N_C^h=N_C^\gamma=1$; there are then $12$ fitting parameters. We typically computed the averaged conductances for system widths $2L = 20, 30, 40, 50, 60, 70$ if the predicted $\honelr$ is less than $1$; otherwise we computed the averaged conductances for  system widths $2L = 20, 30, 40, 50$. By fitting the calculated conductances to the scaling function (\ref{eq:scaling-function}), we have obtained the exponents $\honelr$ and $\galr$ which are reported in Table II of the main paper. We note that the goodness of fit was almost one in all cases. We have obtained the irrelevant exponent $y \approx -1$ for systems with weak finite-size effects. For systems with strong finite-size effects, {\it e.g.} $(\LMon,\RMon)=(1,1),(-2,-2),(-3,-3)$ with $\theta_e=\pi/10$, the irrelevant exponent becomes larger: $y \approx -0.7$ or $-0.8$.

\begin{figure*}[b]
\includegraphics[width=0.9\textwidth]{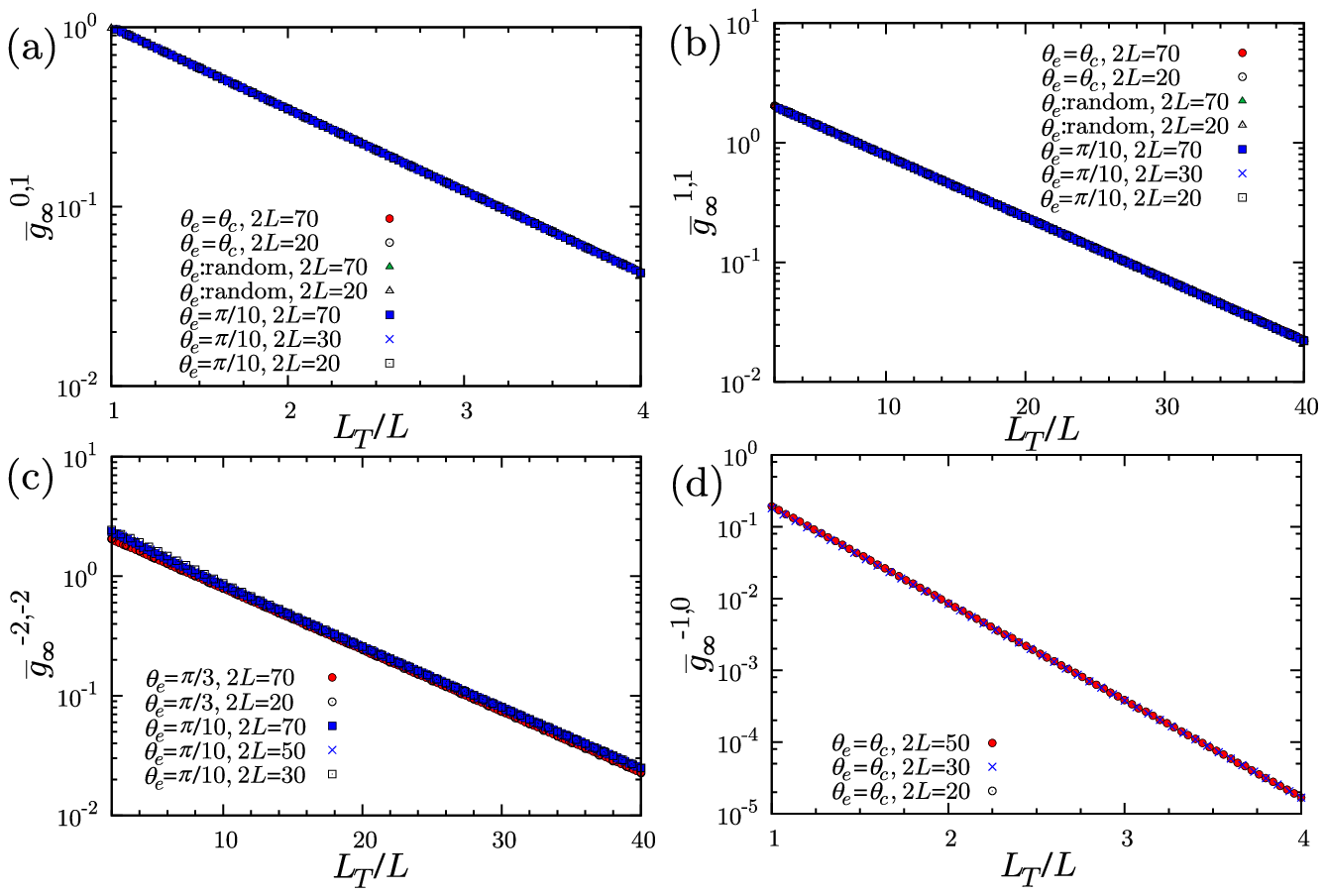} 
\caption{The modified conductance $\bar{g}_\infty^{\LMon,\RMon}$ as a function of $L_T/L$ for systems with different system width $2L$ and $\theta_e$, as
indicated in the figure, in case of (a) $(\LMon,\RMon)=(0,1)$, (b) $(1,1)$, (d) $(-2,-2)$, and (e) $(-1,0)$.  }
\label{fig:conductance_correct}
\end{figure*}

To check the validity of our fitting, we have introduced the modified conductance $\bar{g}_\infty^{\LMon,\RMon}$ defined as
\begin{eqnarray}
\bar{g}_\infty^{\LMon,\RMon} &\equiv& \left(\bar{g}^{\LMon,\RMon} - 2 k^{\LMon,\RMon}_L - \sum_{l=1}^{N_C^h} F_l^h(L,L_T)
 -\sum_{l=0}^{N_C^\gamma} F_l^\gamma(L,L_T) \right) \nonumber \\ &&
/ \exp\left[-\pi\sum_{k=1}^{N^h} a_k^h (2L)^{ky} \frac{L_T}{L}\right],
\end{eqnarray}
so that only the relevant term in the thermodynamic limit is kept in the numerically calculated conductance $\bar{g}^{\LMon,\RMon}$. We have computed $\bar{g}_\infty^{\LMon,\RMon}$ for each width by using the numerically calculated fitting parameters, and results for several cases are presented in Fig. \ref{fig:conductance_correct}. We thus confirm that the modified conductances for all considered widths and $\theta_e$ for each set of $\LMon$ and $\RMon$ collapse well to a single universal exponential function.


\begin{thebibliography}{99}
\bibitem{subleading}
Conformal invariance implies that the exponential corrections to the conductance are given by conformal dimensions in the theory. $\galr$ is then the minimum among $h^{\cal{L},\cal{R}}(2)$ and dimensions of certain primary and descendant fields in the sector $k=1$. Unfortunately, the analysis of \cite{bondesan11} is not sufficient to identify all the fields contributing to $\bar{g}^{{\cal L},{\cal R}}$ in the general case.
\end{thebibliography}
\end{document}